\pdfoutput=1
\documentclass[useAMS,usenatbib,usegraphicx]{mn2e}
\usepackage{apjfonts}
\usepackage{amssymb}
\usepackage{amsmath}
\usepackage{ctable}
\usepackage{fixltx2e} 

\newcommand{\be}{\begin{equation}}
\newcommand{\ee}{\end{equation}}
\newcommand{\ba}{\begin{eqnarray}}
\newcommand{\ea}{\end{eqnarray}}

\newcommand{\Msun}{{\rm M}_{\odot}}
\newcommand{\Zsun}{Z_{\odot}}

\newcommand{\nc}{n_{\rm th}}
\newcommand{\cm}{{\rm cm}}
\newcommand{\pc}{{\rm pc}}

\newcommand{\Myr}{{\rm Myr}}

\newcommand{\fesc}{f_{\rm esc}}
\newcommand{\fion}{\xi_{\rm ion}}
\newcommand{\fions}{\langle\fion\rangle_{\rm single}}
\newcommand{\feff}{f_{\rm esc,\,eff}}
\newcommand{\muv}{{\rm M_{UV}}}
\newcommand{\nstar}{{\rm N_{star}}}
\newcommand{\nuvb}{{\rm N_{UVB}}}
\newcommand{\dd}{{\rm d}}

\title[Binary Stars Can Explain Reionization]
{Binary Stars Can Provide the ``Missing Photons'' Needed for Reionization}

\author[X. Ma et al.]{
  \parbox[t]{1.0\textwidth}{
   Xiangcheng Ma,$^1$\thanks{E-mail: xchma@caltech.edu}
   Philip F. Hopkins,$^1$
   Daniel Kasen,$^{2,3}$
   Eliot Quataert,$^2$
   Claude-Andr{\'e} Faucher-Gigu{\`e}re,$^4$
   Du{\v s}an Kere{\v s}$^5$
   Norman Murray$^6$\thanks{Canada Research Chair in Astrophysics.} and
   Allison Strom$^7$
  }
  \vspace{5pt} \\
  $^1$TAPIR, MC 350-17, California Institute of Technology, Pasadena, CA 91125, USA \\ 
  $^2$Department of Astronomy and Theoretical Astrophysics Center, University of California Berkeley, Berkeley, CA 94720 \\
  $^3$Lawrence Berkeley National Laboratory, 1 Cyclotron Road, Berkeley, CA 94720 \\
  $^4$Department of Physics and Astronomy and CIERA, Northwestern University, 2145 Sheridan Road, Evanston, IL 60208, USA \\
  $^5$Department of Physics, Center for Astrophysics and Space Sciences, University of California at San Diego, 9500 Gilman Drive, La Jolla, CA 92093 \\
  $^6$Canadian Institute for Theoretical Astrophysics, 60 St George Street, University of Toronto, ON M5S 3H8, Canada \\
  $^7$Department of Astrophysics, MC 249-17, California Institute of Technology, Pasadena, CA 91125, USA
}

\pagerange{\pageref{firstpage}--\pageref{lastpage}}
\date{Draft version \today}

\begin{document}
\maketitle
\label{firstpage}

\begin{abstract}
Empirical constraints on reionization require galactic ionizing photon escape fractions $\fesc\gtrsim 20\%$, but recent high-resolution radiation-hydrodynamic calculations have consistently found much lower values $\sim1$--5\%. While these models include strong stellar feedback and additional processes such as runaway stars, they almost exclusively consider stellar evolution models based on single (isolated) stars, despite the fact that most massive stars are in binaries. We re-visit these calculations, combining radiative transfer and high-resolution cosmological simulations from the FIRE project. For the first time, we use a stellar evolution model that includes a physically and observationally motivated treatment of binaries (the BPASS model). Binary mass transfer and mergers enhance the population of massive stars at late times ($\gtrsim 3\,\Myr$) after star formation, which in turn strongly enhances the late-time ionizing photon production (especially at low metallicities). These photons are produced after feedback from massive stars has carved escape channels in the ISM, and so efficiently leak out of galaxies. As a result, the time-averaged ``effective'' escape fraction (ratio of escaped ionizing photons to observed $1500\,$\AA~photons) increases by factors $\sim4$--10, sufficient to explain reionization. While important uncertainties remain, we conclude that binary evolution may be critical for understanding the ionization of the Universe.
\end{abstract}


\section{Introduction}
\label{sec:intro}
The escape fraction ($\fesc$) of hydrogen ionizing photons from high-redshift star-forming galaxies is perhaps the most important and yet most poorly understood parameter in understanding the reionization history. Models of cosmic reionization suggest that $\fesc\gtrsim20\%$ \citep[e.g.][]{kuhlen.faucher.2012:reion,finkelstein.2012:reion.candels,robertson.2013:reion.hudf12,robertson.2015:reion.planck} is needed to match the optical depth of electron scattering inferred from cosmic microwave background (CMB) measurements \citep[e.g.][]{hinshaw.2013:wmap9.cosmo.param,planck.2014:cosmo.param}, assuming that most of the ionizing photons come from star-forming galaxies brighter than $\muv=-13$. 

However, such a high $\fesc$ is problematic in the context of both observations and theory. From the local universe to redshift $z\sim1$, there is no confirmed Lyman continuum (LyC) detection, neither from individual galaxies nor from stacked samples, implying upper limits of $\fesc=1$--3\% \citep[e.g.][]{leitet.2011:fesc.local.haro11,leitet.2013:fesc.local.tol,bridge.2010:fesc.z0pt7.cosmos,siana.2010:fesc.z1pt3.goods}. Even at $z\sim3$, many earlier reports of LyC detection from Lyman break galaxies (LBGs) and Ly$\alpha$ emitters (LAEs) have proven to be contaminated by foreground sources \citep[e.g.][]{siana.2015:lyc.hubble.contam} and a low $\fesc$ about 5\% has been derived from some galaxy samples at this redshift \citep[e.g.][]{iwata.2009:fesc.z3.low,boutsia.2011:fesc.z3pt3.low}.

Moreover, the latest generation of cosmological hydrodynamic simulations predict $\fesc$ to be no more than a few percent in galaxies more massive than $10^9\,\Msun$ in halo mass at $z>6$ \citep[e.g.][]{wise.2014:reion.esc.frac,kimm.cen.2014:esc.frac,paarde.2015:low.esc.frac,ma.2015:fire.escfrac}. These simulations include detailed models of ISM physics, star formation, and stellar feedback, in contrast to early generations of simulations which tended to over-predict $\fesc$ by an order of magnitude, owing to more simplistic ISM models \citep[see][and references therein]{ma.2015:fire.escfrac}. The low $\fesc$ in these simulations is due to the fact that newly formed stars, which dominate the intrinsic ionizing photon budget, begin life buried in their birth clouds, which absorb most of the ionizing photons. By the time low column density escape channels are cleared in the ISM, the massive stars have begun to die and the predicted ionizing photon luminosity has dropped exponentially. Stellar populations older than 3\,Myr have order unity photon escape fractions, but -- according to single stellar evolution models such as {\sc starburst99} \citep{leitherer.1999:sb99} -- these stars only contribute a small fraction of the intrinsic ionizing photon budget \citep{ma.2015:fire.escfrac}. 

Therefore, there appears to be a factor of $\sim4$--5 fewer ionizing photons predicted, compared to what is needed to ionize the Universe. Several solutions have been proposed. For example, \citet{wise.2014:reion.esc.frac} suggested that tiny galaxies that are much fainter than $\muv=-13$ may play a significant role in reionization, since $\fesc$ increases quickly from $5\%$ to order unity for halo mass below $10^{8.5}\,\Msun$. However, others have noted that the required number of tiny galaxies would imply a huge population of Milky Way satellites which have not been observed \citep[see][]{boylan.2014:reion.faint.galaxy,graus.2016:reion.const.lg}. \citet{conroy.kratter.2012:runaway.fesc} proposed that runaway OB stars can boost $\fesc$; however both \citet{kimm.cen.2014:esc.frac} and \citet{ma.2015:fire.escfrac} showed that in high-resolution simulations these produce a marginal effect, increasing $\fesc$ systematically by a factor of only $\sim 1.2$ (far short of the $\gtrsim 4$ required). A more radical alternative is to invoke non-stellar sources for reionization, for example AGN \citep[see e.g.][]{madau.haardt.2015:reion.agn}. This relies on recent observations \citep[e.g.][]{giallongo.2015:agn.lum.func} suggesting much higher number densities of faint AGN at high redshift than previously thought \citep[e.g.][]{hopkins.2007:quasar.lum.func}.

But there are gaps in our understanding of stellar evolution. One key factor that is usually not considered in standard stellar population models is the effect of binary interaction. Mass transfer between binary stars, and binary mergers, can effectively increase the number of high-mass stars at later times after star formation. Also, massive, rapidly-rotating stars produced via mass transfer undergo quasi-homogeneous evolution if the metallicity is sufficiently sub-solar. These stars are hotter and their surface temperature increases as they evolve \citep[e.g.][]{eldridge.stanway.2012:quasi.homo}. All of these can substantially increase the number of ionizing photons produced at late times, compared to what is expected from single-star evolution models \citep[e.g.][]{demink.2014:binary.interaction}. Recently, \citet{stanway.2016:binary.ion.budget} pointed out that the emissivity of ionizing photons from high-redshift galaxies, inferred from their UV luminosities, would be higher by a factor of $\sim1.5$ using stellar evolution models that account for binary interaction. Furthermore, binary evolution does not just produce {\em more} ionizing photons, but it may also substantially increase the escape fractions \citep{ma.2015:fire.escfrac}. 

In this Letter, we explore in more detail the effect of binary interaction on ionizing photon production and escape by repeating the calculation described in \citet{ma.2015:fire.escfrac} using the Binary Population and Spectral Synthesis (BPASS) model of stellar population evolution\footnote{http://bpass.auckland.ac.nz} (\citealt{eldridge.2008:bpass.model}; Eldridge et al, in preparation). These models are calibrated to observations of local stellar populations, and reproduce the observed multiplicity distributions \citep{eldridge.2008:bpass.model}. Moreover, the BPASS model predicts stellar populations with a harder ionizing spectrum, which is required to explain the observed differences between various nebular emission-line properties of metal-poor, younger galaxies at $z\sim2$--3 and local galaxies \citep[][for a more detailed study see Strom et al., in preparation, Steidel et al., in preparation]{steidel.2014:nebular.mosfire}. In \citet{ma.2015:fire.escfrac}, we performed Monte Carlo radiative transfer (MCRT) calculations on a suite of cosmological hydrodynamic simulations and showed that the time-averaged $\fesc$ is about $5\%$ for galaxies of halo masses from $10^9$--$10^{11}\,\Msun$ at $z=6$ using the single-star evolution models from {\sc starburst99}. Importantly, we showed that the results were robust to the resolution of both the radiative transfer calculation and the hydrodynamics (once sufficient resolution for convergence was reached), to variations of the star formation and stellar feedback model, and even to the inclusion of large populations of runaway stars. We will show here, however, that the inclusion of binary evolution effects increases the predicted escape fractions substantially, reconciling them with constraints on reionization. We describe the simulation and radiative transfer code in Section \ref{sec:method}, present the results in Section \ref{sec:results}, and conclude in Section \ref{sec:con}.

We adopt a standard flat $\Lambda$CDM cosmology with cosmological parameters $H_0=70.2 {\rm~km~s^{-1}~Mpc^{-1}}$, $\Omega_{\Lambda}=0.728$, $\Omega_{m}=1-\Omega_{\Lambda}=0.272$, $\Omega_b=0.0455$, $\sigma_8=0.807$ and $n=0.961$, consistent with observations \citep[e.g.][]{hinshaw.2013:wmap9.cosmo.param,planck.2014:cosmo.param}.

\begin{table}
\begin{center}
\caption{Simulations analyzed in this paper.} 
\label{tbl:sim}
\begin{tabular}{lcccccccc}
\hline\hline
Name & $m_b$ & $\epsilon_b$ & $m_{\rm dm}$ & $\epsilon_{\rm dm}$ & $M_{\rm vir}$ & $M_{\ast}$ & $\rm M_{UV}$ \\
 & ($\Msun$) & ($\pc$) & ($\Msun$) & ($\pc$) & ($\Msun$) & ($\Msun$) & (AB mag) \\
\hline
z5m09 & 16.8 & 0.14 & 81.9 & 5.6 & 7.6e8 & 3.1e5 & -10.1 \\
z5m10mr & 1.1e3 & 1.9 & 5.2e3 & 14 & 1.5e10 & 5.0e7 & -17.5 \\
z5m11 & 2.1e3 & 4.2 & 1.0e4 & 14 & 5.6e10 & 2.0e8 & -18.5 \\
\hline \\
\multicolumn{8}{p{\linewidth}}{{\bf Notes.} Initial conditions and galaxy properties at $z=6$.} \\
\multicolumn{8}{p{\linewidth}}{(1) Name: Simulation designation.} \\
\multicolumn{8}{p{\linewidth}}{(2) $m_b$: Initial baryonic particle mass.} \\
\multicolumn{8}{p{\linewidth}}{(3) $\epsilon_b$: Minimum baryonic force softening. Force softening is adaptive.} \\
\multicolumn{8}{p{\linewidth}}{(4) $m_{\rm dm}$: Dark matter particle mass in the high-resolution regions.} \\
\multicolumn{8}{p{\linewidth}}{(5) $\epsilon_{\rm dm}$: Minimum dark matter force softening.} \\
\multicolumn{8}{p{\linewidth}}{(6) $M_{\rm vir}$: Halo mass of the primary galaxy at $z=6$.} \\
\multicolumn{8}{p{\linewidth}}{(7) $M_{\ast}$: Stellar mass of the primary galaxy at $z=6$.} \\
\multicolumn{8}{p{\linewidth}}{(8) $\rm M_{UV}$: Galaxy UV magnitude (absolute AB magnitude at 1500 \AA).} \\
\end{tabular}
\end{center}
\end{table}

\section{Method}
\label{sec:method}
In this work, we study the effect of binary evolution on $\fesc$ using three galaxies from a suite of cosmological zoom-in simulations presented in \citet{ma.2015:fire.escfrac}. The simulation and radiative transfer are identical. We {\em only} replace the stellar evolution model used for the post-processing radiative transfer calculations. This is likely to result in a lower limit on the impact of binaries on $\fesc$, because we do not include the enhanced radiative feedback due to binaries in our simulation. We briefly review the methodology here, but refer to \citet{ma.2015:fire.escfrac} for more details.

The simulations are part of the Feedback in Realistic Environment project\footnote{http://fire.northwestern.edu} \citep[FIRE;][]{hopkins.2014:fire.galaxy}. They are run using {\sc gizmo} \citep{hopkins.2015:gizmo.code} in P-SPH mode, which adopts a Lagrangian pressure-entropy formulation of the smoothed particle hydrodynamics (SPH) equations that improves the treatment of fluid-mixing instabilities \citep{hopkins.2013:psph.code}. Galaxy properties at $z=6$ for the three simulations used in this work (z5m09, z5m10mr, and z5m11) are listed in Table \ref{tbl:sim}. Our simulations span halo masses from $10^9$--$10^{11}\,\Msun$ at $z=6$. These galaxies lie on the low-mass extrapolations of the observed stellar mass--halo mass relation and SFR--stellar mass relation at $z>6$ \citep{ma.2015:fire.escfrac}. At lower redshifts (where observations exist at these masses), the simulations have been shown to reproduce observed scaling relations and chemical abundances \citep{hopkins.2014:fire.galaxy,ma.2016:fire.mzr}, properties of galactic outflows and circum-galactic absorbers \citep{muratov.2015:fire.mass.loading,faucher.2015:fire.neutral.hydrogen,faucher.2016:fire.neutral.hydrogen}, and abundances and kinematics of observed (local) dwarfs in this mass range \citep{onorbe.2015:fire.dwarf,chan.2015:fire.cusp.core}.

In the simulations, gas follows an ionized-atomic-molecular cooling curve from $10$--$10^{10}$\,K, including metallicity-dependent fine-structure and molecular cooling at low temperatures and high-temperature metal-line cooling for 11 separately tracked species \citep{wiersma.2009:metal.cooling}. We do not include a primordial chemistry network nor consider Pop III star formation, but apply a metallicity floor of $Z=10^{-4}\,\Zsun$. At each timestep, the ionization states are determined following \citet{katz.1996:sph.cooling} and cooling rates are computed from a compilation of {\sc cloudy} runs, including a uniform but redshift-dependent photo-ionizing background tabulated in \citet{faucher.2009:uvb}, and an approximate model of photo-ionizing and photo-electric heating from local sources. Gas self-shielding is accounted for with a local Jeans-length approximation, which is consistent with the radiative transfer calculations in \citet{faucher.2010:lya.cooling}. The on-the-fly calculation of ionization states is consistent with more accurate post-processing radiative transfer calculations \citep{ma.2015:fire.escfrac}.

We follow the star formation criteria in \citet{hopkins.2013:sf.criteria} and allow star formation to take place only in dense, molecular, and self-gravitating regions with hydrogen number density above a threshold $\nc=50~\cm^{-3}$. Stars form at 100\% efficiency per free-fall time when the gas meets these criteria, and there is no star formation elsewhere. Because we require star-forming gas to be self-gravitating, its effective density is even higher than the fiducial density threshold we adopt in our simulations. We emphasize the importance of resolving the formation and destruction of individual star-forming regions in accurately predicting $\fesc$, as stressed also in other studies \citep[e.g.][]{kimm.cen.2014:esc.frac,paarde.2015:low.esc.frac,ma.2015:fire.escfrac}. Simulations using unphysically low $\nc$ fail to resolve this and tend to over-predict $\fesc$ by an order of magnitude \citep[see][and reference therein]{ma.2015:fire.escfrac}.

The simulations include several different stellar feedback mechanisms, including (1) local and long-range momentum flux from radiative pressure, (2) energy, momentum, mass and metal injection from SNe and stellar winds, and (3) photo-ionization and photo-electric heating. We follow \citet{wiersma.2009:chemical.enrich} and include metal production from Type-II SNe, Type-Ia SNe, and stellar winds. Every star particle is treated as a single stellar population with known mass, age, and metallicity, assuming a \citet{kroupa.2002:imf} initial mass function (IMF) from $0.1$--$100~\Msun$. The feedback strengths are directly tabulated from {\sc starburst99}.


For every snapshot, we map the main galaxy onto a Cartesian grid of side length $L$ equal to two virial radii and with $N$ cells along each dimension. We choose $N=256$ for z5m09 and z5m10mr and $N=300$ for z5m11, so that the cell size $l=L/N$ varies but is always smaller than 100 pc. This ensures convergence of the MCRT calculation \citep{ma.2015:fire.escfrac}. The MCRT code we use is derived from the MCRT code {\sc sedona} \citep{kasen.2006:sedona.code}, but focuses on radiative transfer of hydrogen ionizing photons. The MCRT method is similar to that described in \citet{fumagalli.2011:rt.line.transfer,fumagalli.2014:rt.cgm.abs}. $\nstar=3\times10^7$ photon packets are isotropically emitted from the location of star particles, sampling their ionizing photon budgets. Another $\nuvb=3\times10^7$ photon packets are emitted from the boundary of the computational domain in a manner that produces a uniform, isotropic ionizing background with intensity given by \citet{faucher.2009:uvb}. The MCRT code includes photoionization, collisional ionization, recombination, and dust absorption and uses an iterative method to reach photoionization equilibrium. The numbers of photon packets and iteration are selected to ensure convergence.

\begin{figure}
\centering
\includegraphics[width=0.5\textwidth]{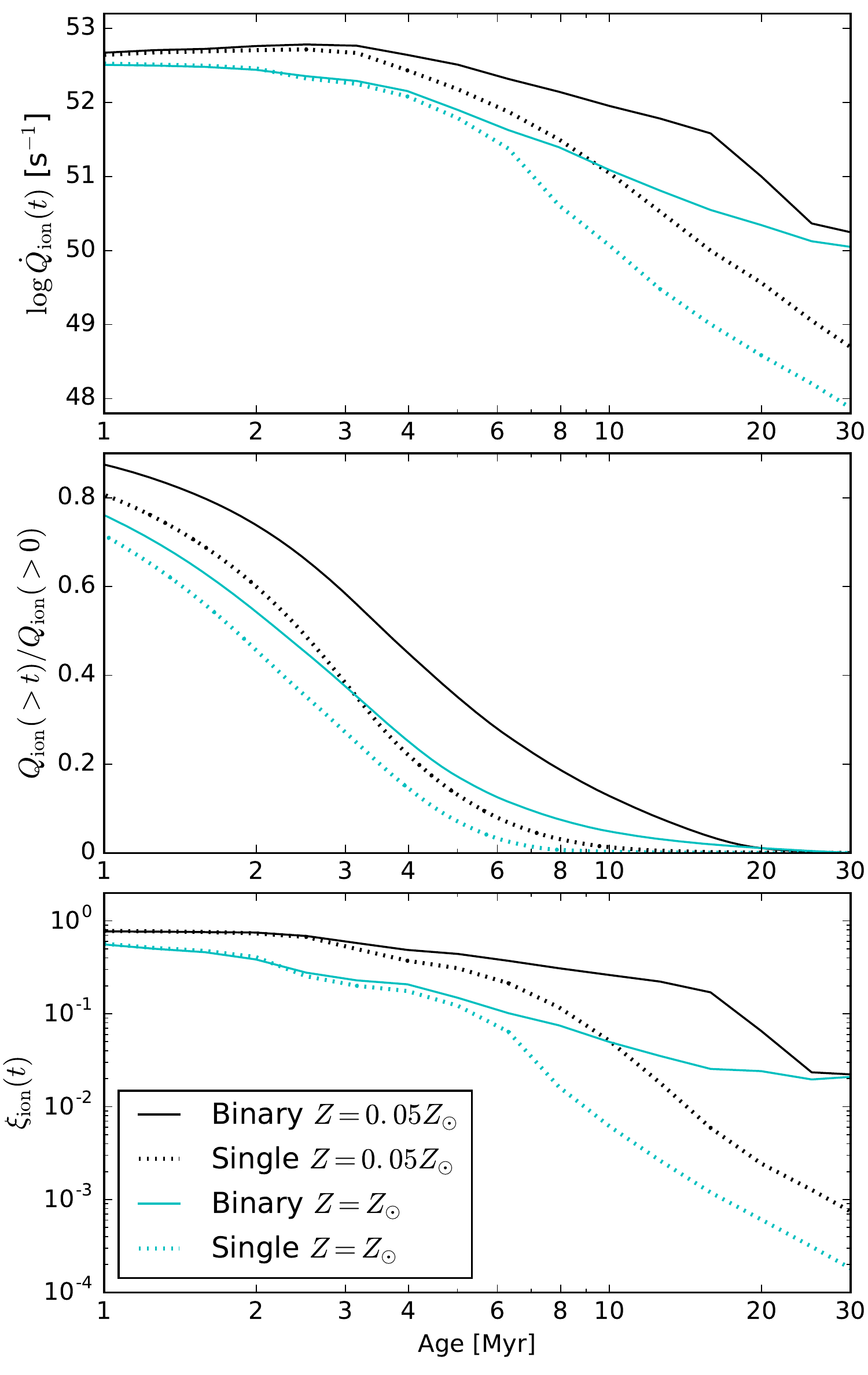}
\caption{{\em Top:} Ionizing photon production rate, $\dot{Q}_{\rm ion}(t)$ as a function of age for a $10^6~\Msun$ star cluster, predicted by different stellar evolution models. {\em Middle:} Fraction of ionizing photons emitted after time $t$. {\em Bottom:} Ratio of ionizing luminosity to $1500$ \AA~luminosity, $\fion$ as a function of age for the same star cluster. We show both single-star models (dotted) and binary models (solid) at metallicities $Z=0.05~\Zsun$ (black) and $Z=\Zsun$ (cyan), respectively. 
Including binaries leads to more massive stars at late times (from mass transfer and mergers), which dramatically enhances the ionizing photon production after $t\sim3$\,Myr. About 60\% (20\%) of the ionizing photons are emitted after 3\,Myr (10\,Myr) in the low-metallicity binary model, while this fraction is much lower in single-star model as well as at solar metallicity. The difference between single-star and binary models is less significant at solar metallicity. {\sc starburst99} models, which also ignore binaries, are nearly identical to the BPASS single-star models at both metallicities.}
\label{fig:spectra}
\end{figure}

\begin{figure}
\centering
\includegraphics[width=0.5\textwidth]{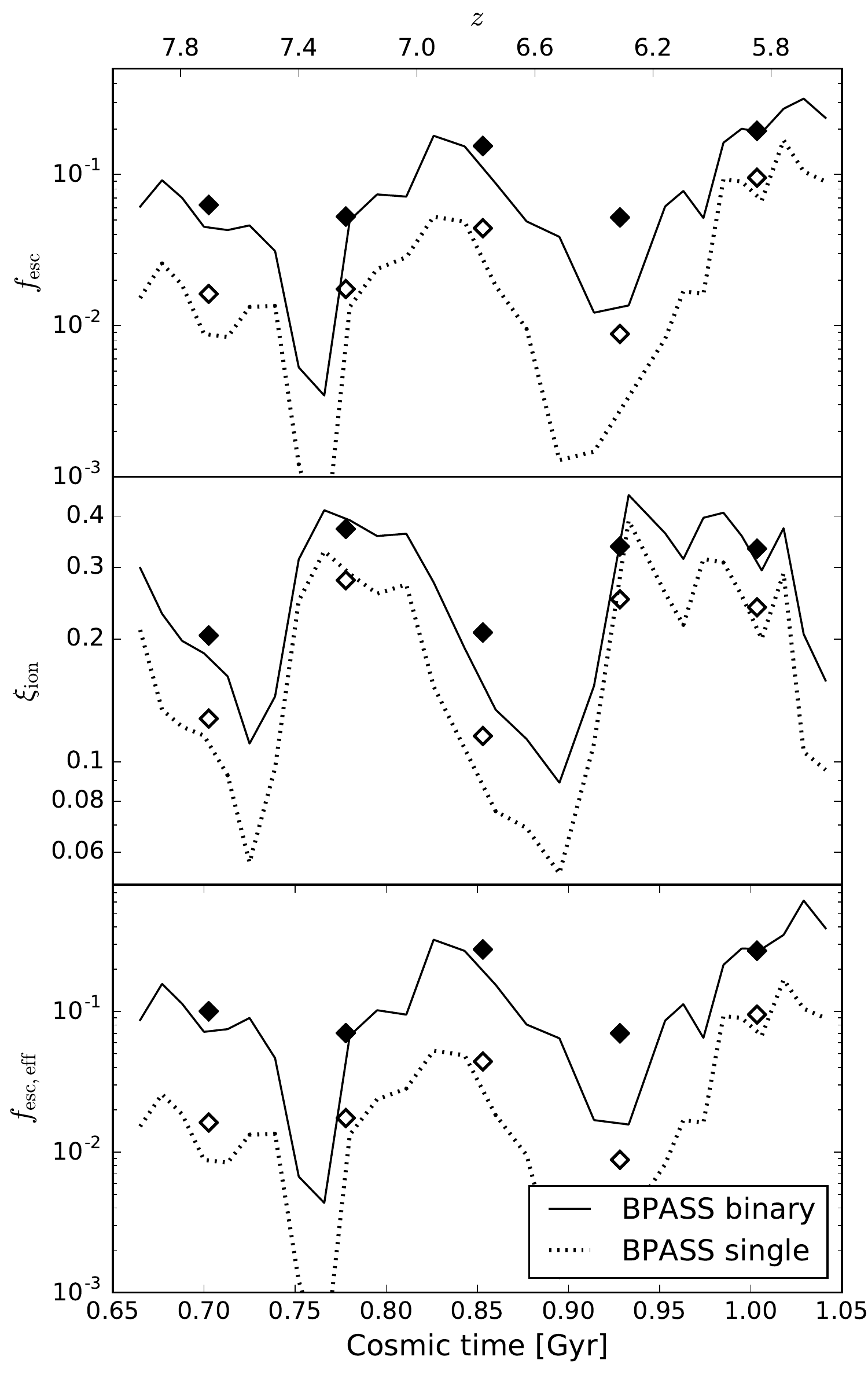}
\caption{{\em Top:} True ionizing photon escape fraction $\fesc$, in our z5m10mr simulation (a $\sim 10^{10}~\Msun$ halo at $z=6$) as a function of redshift (or cosmic time). Lines show the instantaneous values in each snapshot, symbols are time-averaged in $100$ Myr intervals. {\em Middle:} $\fion$ (as Fig.~\ref{fig:spectra}) as a function of time. {\em Bottom:} Effective escape fraction, $\feff$ (Equation \ref{eqn:feff}) as a function of time. In the single-star model, $\fesc\lesssim 5\%$ most of the time, insufficient for reionization. Accounting for binary effects boosts $\fion$ by a factor $\sim 1.5$ -- considerable but insufficient to explain reionization. But it also boosts $\fesc$ by factors $\sim3$--6 because the ionizing photons produced later (after $t\gtrsim 3\,$Myr) preferentially escape, so the ``effective escape fraction'' $\feff$ is increased by factors $\sim4$--10 and reaches the $\sim 20\%$ values needed to explain reionization.}
\label{fig:rt}
\end{figure}

\begin{figure}
\centering
\includegraphics[width=0.5\textwidth]{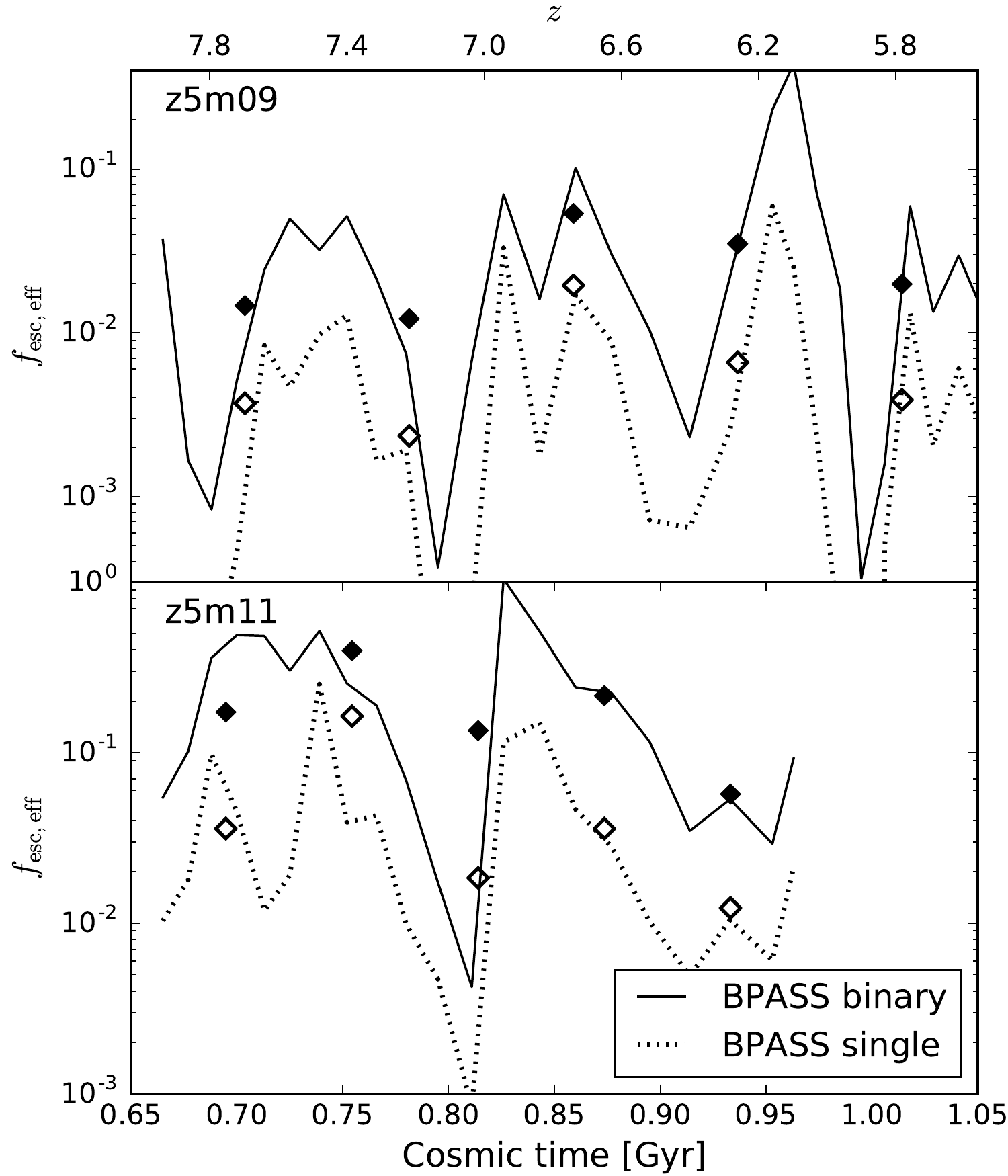}
\caption{Effective escape fraction as a function time for z5m09 and z5m11. Like z5m10 (Figure \ref{fig:rt}), binary stellar models boost $\feff$ by factors of $\sim4$--10. In more massive galaxies, the mean $\feff$ reaches $\sim20\%$, sufficient for reionization.}
\label{fig:z5}
\end{figure}

\section{Results}
\label{sec:results}
We use $\dot{Q}_{\rm ion}(t)$ to represent the ionizing photon production rate (number of ionizing photons produced per second) and
\be
Q_{\rm ion}(>t) = \int_t^\infty \dot{Q}_{\rm ion}(t') \, \dd t'
\ee
to represent the number of ionizing photons produced after time t. In Figure \ref{fig:spectra}, we show $\dot{Q}_{\rm ion}(t)$ (upper panel), the {\em fraction} of ionizing photons emitted after time $t$, $Q_{\rm ion}(>t)/Q_{\rm ion}(>0)$ (middle panel), and the ratio between ionizing luminosity and the luminosity at $1500$\,\AA,
\be
\label{eqn:fion}
\fion = \frac{\int_{10\,\text{\AA}}^{912\,\text{\AA}} L_{\lambda} \dd\lambda}{\lambda L_{\lambda}(1500\,\text{\AA})}
\ee
(bottom panel), as a function of age, of an instantaneously formed star cluster of mass $10^6\,\Msun$, for several stellar population models from BPASS. We adopt a \citet{kroupa.2002:imf} IMF with slopes of $-1.3$ from 0.1--$0.5~\Msun$ and $-2.35$ from 0.5--$100\,\Msun$, consistent with that used in the simulation. We show the BPASS model at metallicity $Z=0.001$ ($Z=0.05\,\Zsun$, black), the lowest metallicity available and the closest to our simulations, for both single-star evolution (dotted) and binary evolution (solid) models. We also compare those with $Z=0.02$ ($Z=\Zsun$, cyan) models from BPASS. We note that the STARBURST99 models (not shown), which are the default model in \citet{ma.2015:fire.escfrac}, are nearly identical to the single-star model from BPASS at both metallicities.

The ionizing photon production rates in the single-star and binary models are very similar for the first 3\,Myr, but start to differ significantly after 3\,Myr at $Z=0.05\,\Zsun$, with the binary model producing an order of magnitude more ionizing photons by 10\,Myr. Also, in the binary model, the production of ionizing photons is more extended. For example, almost 60\% (20\%) of the ionizing photons are produced after 3\,Myr (10\,Myr), while this fraction is 40\% (1\%) in single-star model. These late-time photons can escape more easily so one should expect them to make a big difference on $\fesc$ (as confirmed by MCRT calculations below). However, at solar metallicity, these fractions are much lower and the difference between single-star and binary models is less significant.

To illustrate the effects of binaries, we run our MCRT code on the main galaxy in our z5m10mr simulation (a $\sim10^{10}\,\Msun$ halo at $z=6$) and compute $\fesc$ using both single-star and binary BPASS models with $Z=0.05\,\Zsun$. The results are presented in Figure \ref{fig:rt}. Lines and symbols show the instantaneous value and time-averaged  values over $\sim100\,\Myr$, respectively. Dotted lines and open symbols represent the single-star model, while solid lines and filled symbols represent the binary model. From top to bottom, the three panels show $\fesc$ (the ``true'' fraction of ionizing photons that escape the galaxy virial radius), $\fion$, and the ``effective'' escape fraction from $z=5.5$--8. The effective escape fraction is defined as
\be
\label{eqn:feff}
\feff = \fesc \frac{\fion}{\fions},
\ee
which is the ratio of the escaping ionizing flux to 1500 \AA~flux, relative to what would be computed using single-star models. $\feff$ simply equals $\fesc$ for single-star models, while for binary models, it also accounts for the change of $\fion$ relative to single-star models. 

The instantaneous $\fesc$ is highly time-variable, associated with stochastic formation and destruction of individual star-forming clouds (consistent with several other studies \citealt{wise.2014:reion.esc.frac,kimm.cen.2014:esc.frac,paarde.2015:low.esc.frac}). For single-star models, $\fesc$ is below $5\%$ most of the time, because young stars are buried in their birth clouds, which prevent almost all ionizing photons from escaping. Most of the photons that escape come from stellar populations with age $\sim3$--10\,Myr, but they only contribute a very small fraction of the intrinsic ionizing photons in single-star models \citep{ma.2015:fire.escfrac}. However, at all times, the binary model predicts significantly higher (factors $\sim3$--6) values for $\fesc$. We also find that $\fion$ is boosted by a factor of $\sim1.5$, consistent with \citet{stanway.2016:binary.ion.budget}. Multiplying the two factors, we find that the effective escape fraction is boosted by factors of $\sim4$--10, with most of the contribution coming from the increased $\fesc$. Averaged over the entire redshift range $z=5.5$--8 that we consider here, accounting for binary effects increases the true ionizing escape fraction $\fesc$ from $6\%$ to $14\%$ and increases $\feff$ from $6\%$ to $\sim 20\%$. This is consistent with what is required in empirical reionization models.

In Figure \ref{fig:z5}, we show the effective escape fraction as a function of time for z5m09 and z5m11 galaxies. Similar to z5m10mr (Figure \ref{fig:rt}), binary models boost time-averaged $\feff$ by factors of $\sim 4$--10. In the lowest mass halo (z5m09), $\feff$ is still low, because halo gas is largely neutral close to the galaxy in such low-mass systems, which consumes a large fraction of the ionizing photons. In more massive galaxies like z5m10mr and z5m11, $\feff$ can reach $\gtrsim20\%$.

\section{Discussion and Conclusions}
\label{sec:con}
In this work, we study the effect of binary evolution on ionizing photon production and escape in high-redshift galaxies, using three high-resolution cosmological simulations from the FIRE project. The simulated galaxies are around the mass estimated to dominate re-ionization (halo $M_{\rm halo}=10^9$--$10^{11}\,\Msun$ at $z=6$). Using detailed radiative transfer calculations, we show that recent stellar evolution models which account for mass transfer and mergers in binaries (specifically, the BPASS model) produce significantly more ionizing photons for stellar populations older than 3\,Myr compared to stellar evolution models ignoring binaries. These later-time photons easily escape, collectively increasing the escape fraction and ionizing photon production rate dramatically from high-redshift low-metallicity galaxies.

For single-star evolution models, we predict $\fesc$ below $5\%$ most of the time, less than what is required ($\sim20\%$) for cosmic reionization. However, when accounting for binary effects, $\fesc$ can be boosted by factors of $\sim3$--6 and $\fion$ can be boosted by a factor of $1.5$. Therefore, the ``effective'' escape fraction (the ratio of escaped ionizing photon flux to 1500\,\AA~flux) can be boosted by factors of $\sim4$--10. For the more massive galaxies in our simulation, this brings them into good agreement with the values required to ionize the Universe. 

We emphasize that the most important change relative to single-star models is not in the absolute photon production rate, but its time-dependence, because photons emitted after 3\,Myr can much more easily escape star-forming complexes once feedback from massive stars has destroyed the dense birth cloud. Moreover, we have exhaustively tested in a previous study \citep{ma.2015:fire.escfrac} that these results are not sensitive to our star formation and stellar feedback models. For example, increasing the strength of all feedback (radiation, stellar winds, SNe) per star particle relative to our fiducial model simply leads to self-regulation at lower star formation rates, giving an identical prediction for $\fesc$. Likewise, increasing the density threshold for star formation, re-distributing ionizing photons to fewer but more luminous particles, increasing the ionizing photon production rate used for feedback in the code, all produce similar predictions for $\fesc$.

Nevertheless, the binary fraction in high-redshift galaxies and the details of binary evolution are both uncertain, so our results are not definitive. They do, however, demonstrate the potential for binary evolution to reconcile empirical constraints on reionization by starlight with observations and simulations. In principle, these models can be confronted by the observed GRB rates at these redshifts \citep[e.g.][]{kistler.2009:grb.highz.sfr,wyithe.2010:grb.highz.esc.frac}, although large uncertainties remain. In addition, the BPASS model does not include stellar rotation before binary interaction, which may also significantly increase the intrinsic ionizing photon production rate \citep[e.g.][]{topping.shull.2015:rotation.ion}. Rotation likely has a smaller effect on $\fesc$ because most of the extra ionizing photons it predicts are produced less than 3\,Myr after star formation.

We have repeated our radiative transfer calculation on cosmological simulations of Milky Way-mass galaxies at $z=0$ ($Z\sim\Zsun$) from the FIRE project \citep[see][for details]{hopkins.2014:fire.galaxy}. We find that binaries appear to be enhancing $\fesc$ by only a factor $\sim 1.5$ at solar metallicity. This is expected since binary effects tend to be weaker at higher metallicities (also see Figure \ref{fig:spectra}), for at least three reasons: (1) the number of ionizing photons decreases significantly as stellar atmospheres are cooler, (2) quasi-homogenous evolution ceases to apply above $Z=0.004$ ($Z=0.2\,\Zsun$), and (3) stellar winds become stronger, reducing the lifetime of massive stars and suppressing the mass transfered between binaries. In addition, the absolute time-averaged $\fesc$ does not exceed $\sim 3\%$ in these galaxies, consistent with observational constraints in the local Universe (see references in Section \ref{sec:intro}). It appears that these galaxies are forming stars in a more ``calm'', less-bursty mode compared to the high-redshift dwarfs here \citep[e.g.][]{sparre.2015:fire.sf.burst}, and maintain much larger reservoirs of neutral gas in their galactic disks, which leads to much larger absorption even of photons produced by intermediate-age massive stars. A detailed study will be presented in a separate paper (Su et al., in preparation).

\section*{Acknowledgments}
We thank Chuck Steidel for helpful discussions and the referee for useful comments. We also thank John Beacom and Mike Shull for helpful suggestions after the paper was submitted to arXiv.
The simulations used in this paper were run on XSEDE computational resources (allocations TG-AST120025, TG-AST130039, and TG-AST140023). 
The analysis was performed on the Caltech compute cluster ``Zwicky'' (NSF MRI award \#PHY-0960291).
Support for PFH was provided by an Alfred P. Sloan Research Fellowship, NASA ATP Grant NNX14AH35G, and NSF Collaborative Research Grant \#1411920 and CAREER grant \#1455342.
D. Kasen is supported in part by a Department of Energy Office of Nuclear Physics Early Career Award, and by the Director, Office of Energy Research, Office of High Energy and Nuclear Physics, Divisions of Nuclear Physics, of the U.S. Department of Energy under Contract No. DE-AC02-05CH11231 and by the NSF through grant AST-1109896. 
D. Kere{\v s} was supported by NSF grant AST-1412153 and funds from the University of California, San Diego. 
CAFG was supported by NSF through grants AST-1412836 and AST-1517491, by NASA through grant NNX15AB22G, and by STScI through grant HST-AR-14293.001-A.
EQ was supported by NASA ATP grant 12-APT12-0183, a Simons Investigator award from the Simons Foundation, and the David and Lucile Packard Foundation.

\bibliography{ms}

\label{lastpage}

\end{document}